%% file: draft_Dp_omegamuv_arXiv_v2.tex
\documentclass[aps,prd,twocolumn,showpacs,amsmath,amssymb]{revtex4-1}
\usepackage{amsmath}
\usepackage{graphicx}
\usepackage{subfigure}
\usepackage{epstopdf}
\usepackage{color}
\usepackage{multirow}
\usepackage{setspace}
\usepackage{overpic}
\usepackage{amssymb}
\usepackage[dvipdfmx, bookmarksnumbered, pdfstartview=FitH,colorlinks,urlcolor=blue, citecolor=blue,linkcolor=blue] {hyperref}
\usepackage{lineno}
\usepackage{bm}
\usepackage{rotating}
\usepackage[utf8]{inputenc}

\let\oldequation\equation
 \let\oldendequation\endequation
 \renewenvironment{equation}
   {\linenomathNonumbers\oldequation}
   {\oldendequation\endlinenomath}

\begin{document}

\title{\bf \boldmath
Observation of the semi-muonic decay $D^+\to \omega\mu^+\nu_\mu$
}

\input{authorlist_CWR_27Nov2019}

\begin{abstract}
  We report the first observation
  of the semi-muonic decay $D^+\to \omega \mu^+\nu_\mu$
  using an $e^+e^-$ collision data sample corresponding to an integrated luminosity of
  $2.93~\mathrm{fb}^{-1}$ collected with the BESIII detector at a
  center-of-mass energy of 3.773 GeV.
  The absolute branching fraction of the $D^+\to \omega \mu^+\nu_\mu$ decay is
  measured to be ${\mathcal B}_{D^+\to \omega
    \mu^+\nu_\mu}=(17.7\pm1.8_{\rm stat}\pm1.1_{\rm syst})\times
  10^{-4}$. Its ratio with the world average value of the branching fraction of
  the $D^+\to \omega e^+\nu_e$ decay probes lepton flavor universality and it
  is determined to be
  ${\mathcal B}_{D^+\to \omega \mu^+\nu_\mu}/{\mathcal B}^{\rm
    PDG}_{D^+\to \omega e^+\nu_e}=1.05\pm0.14$,
  in agreement with the standard model expectation
  within one standard deviation.
\end{abstract}


\maketitle

\oddsidemargin  -0.2cm
\evensidemargin -0.2cm

Lepton flavor universality (LFU) is one of key predictions in the standard model.
It requires that the couplings between three generation leptons and gauge bosons are equal with each other.
Studies of the semi-leptonic (SL) decays of pseudoscalar mesons
are powerful to test LFU.
In recent years, the difference between the branching fraction (BF) ratio ${\mathcal
  R}_{\tau/\ell}={\mathcal B}_{B\to \bar
  D^{(*)}\tau^+\nu_\tau}/{\mathcal B}_{B\to \bar
  D^{(*)}\ell^+\nu_\ell}$~($\ell=\mu$,
$e$)~\cite{babar_1,babar_2,lhcb_1,belle2015,belle2016,lhcb_1a,belle2019}
and the standard model predictions is found to be larger than three standard deviations~\cite{hflav2018}.
Possible physics mechanisms~\cite{Fajfer,Branco} were proposed to explain this tension.
Comprehensive tests of $e$-$\mu$ LFU with SL $D$ decays, especially for those lesser-known decays, offer critical information for thorough exploration of LFU.

In 2018, BESIII
reported tests of LFU with $D\to \pi \ell^+\nu_\ell$ decays~\cite{bes3-pimuv} which are mediated via $c\to d\ell^+\nu_\ell$.
The difference between the BF ratio ${\mathcal R}_{\mu/e}^{c\to d}={\mathcal B}_{D\to \pi\mu^+\nu_\mu}/{\mathcal B}_{D\to \pi e^+\nu_e}$~\cite{pdg2018}
and the SM prediction is found to be greater than one standard deviation.
Tests of LFU with other SL $D$ decays mediated via $c\to d\ell^+\nu_\ell$ are important to understand this situation.
One possible candidate decay is $D^+\to\omega\mu^+\nu_\mu$. Although this decay was theoretically predicted before 1990~\cite{isgw},
it has never been experimentally confirmed yet to date.
Since 2015, different theoretical approaches, e.g., light-front quark model (LFQM)~\cite{cheng}, re-called chiral
  unitary approach ($\chi$UA)~\cite{Sekihara}, covariant confined
  quark model (CCQM)~\cite{soni1}, light-cone QCD sum rules
  (LCSR)~\cite{soni2,lcsr2}, and relativistic quark model~(RQM)~\cite{rqm},
were adopted to investigate $D^+\to \omega \mu^+\nu_\mu$.
The predicted BFs range between (1.78-2.46)$\times 10^{-3}$~\cite{cheng,Sekihara,soni1,soni2,lcsr2,rqm}, 
corresponding to the BF ratios ${\mathcal
  B}_{D^+\to \omega \mu^+\nu_\mu}/{\mathcal B}_{D^+\to
  \omega e^+\nu_e}$ of (0.93-0.99).
Observation and measurement of the BF of $D^+\to \omega \mu^+\nu_\mu$ are crucial to test $e$-$\mu$ LFU
with $D^+\to \omega\ell^+\nu_\ell$ decays.
The measured BF are also important to distinguish between various theoretical calculations,
thereby improving understanding of nonperturbative effects in heavy meson decays~\cite{isgw2,nonper}.

This paper reports the first observation and BF measurement of $D^+\to \omega
\mu^+\nu_\mu$ as well as a test of $e$-$\mu$ LFU with $D^+\to \omega\ell^+\nu_\ell$ decays,
by analyzing 2.93~fb$^{-1}$ of data accumulated with the
  BESIII detector at a center-of-mass energy $\sqrt s=$
  3.773~GeV~\cite{lum}.
Throughout this paper, charge conjugated channels are implied.

  The BESIII detector is a magnetic spectrometer~\cite{BESCol} located
  at the Beijing Electron Positron Collider
  (BEPCII)~\cite{BEPCII}. The cylindrical core of the BESIII detector
  consists of a helium-based main drift chamber (MDC), a plastic
  scintillator time-of-flight system (TOF), and a CsI(Tl)
  electromagnetic calorimeter (EMC), which are all enclosed in a
  superconducting solenoidal magnet providing a 1.0 T magnetic
  field. The solenoid is supported by an octagonal flux-return yoke
  with resistive plate counter muon identifier modules interleaved
  with steel. The acceptance of charged particles and photons is 93\%
  over $4\pi$ solid angle. At 1 GeV/$c$, the charged-particle momentum
  resolution is 0.5\%, and the $dE/dx$ resolution is 6\% for electrons
  from Bhabha scattering. The EMC measures photon energies with a
  resolution of 2.5\% (5\%) at 1 GeV in the barrel (end-cap)
  region. The time resolution of the TOF barrel part is 68~ps, while
  that of the end-cap part is 110 ps. More details about the BESIII
  detector are described in Ref.~\cite{BESCol}.

  Simulated samples produced with the {\sc geant4}-based~\cite{geant4}
  Monte Carlo (MC) software, which includes the geometric description~
  \cite{geod1,geod2} of the BESIII detector and the detector response,
  are used to determine the detection efficiency and to estimate the
  backgrounds. The simulation includes the beam-energy spread and
  initial-state radiation (ISR) in the $e^+e^-$ annihilations modeled
  with the generator {\sc kkmc}~\cite{ref:kkmc}.  The inclusive MC
  samples consist of the production of the $D\bar{D}$ pairs, the
  non-$D\bar{D}$ decays of the $\psi(3770)$, the ISR production of the
  $J/\psi$ and $\psi(3686)$ states, and the continuum processes
  ($e^+e^-\to q\bar q$, ($q=u,d,s$)) incorporated in {\sc
    kkmc}~\cite{ref:kkmc}.  The known decay modes are modeled with
  {\sc evtgen}~\cite{ref:evtgen} using BFs taken from the Particle
  Data Group (PDG)~\cite{pdg2018}, and the remaining unknown decays
  from the charmonium states with {\sc
    lundcharm}~\cite{ref:lundcharm}. The final-state radiation from
  charged final-state particles is incorporated with the {\sc photos}
  package~\cite{photos}.
  The $D^+\to \omega\mu^+\nu_\mu$ decay is
  simulated by a model with the form factor parameters of
  $r_V=V(0)/A_1(0)=1.24\pm0.11$ and $r_2=A_2(0)/A_1(0)=1.06\pm0.16$, which are quoted from
  Ref.~\cite{BESIII-omegaev}.

  At $\sqrt s=3.773$ GeV, the $\psi$(3770) resonance decays
  predominately into $D^0\bar{D}^0$ or $D^+D^-$ meson pairs.  The
  $D^-$ mesons are reconstructed by their hadronic decays to
  $K^{+}\pi^{-}\pi^{-}$, $K^0_{S}\pi^{-}$,
  $K^{+}\pi^{-}\pi^{-}\pi^{0}$, $K^0_{S}\pi^{-}\pi^{0}$,
  $K^0_{S}\pi^{+}\pi^{-}\pi^{-}$, and $K^{+}K^{-}\pi^{-}$, and
  referred to as single-tag (ST) $D^-$ mesons.  In the sides recoiling
  against of the ST $D^-$ mesons, the candidate $D^+\to \omega \mu^+
  \nu_\mu$ decays are selected to form double-tag (DT) events.  The
  absolute BF of $D^+\to \omega \mu^+ \nu_\mu$ is determined by
\begin{equation}
\label{eq:bf}
{\mathcal B}_{\rm SL}=N_{\mathrm{DT}}/(N_{\mathrm{ST}}^{\rm tot}\cdot \varepsilon_{\rm SL}\cdot{\mathcal B}_{\omega}\cdot{\mathcal B}_{\pi^0}),
\end{equation}
  where $N_{\rm ST}^{\rm tot}$ and $N_{\rm DT}$ are the ST and DT
  yields in the data sample, ${\mathcal B}_{\omega}$ and ${\mathcal
    B}_{\pi^0}$ are the BFs of the $\omega\to\pi^+\pi^-\pi^0$ and
  $\pi^0\to\gamma\gamma$ decays, respectively, and $\varepsilon_{\rm
    SL}=\Sigma_i [(\varepsilon^i_{\rm DT}\cdot N^i_{\rm
    ST})/(\varepsilon^i_{\rm ST}\cdot N^{\rm tot}_{\rm ST})]$ is the
  efficiency of detecting the SL decay in the presence of the ST $D^-$
  meson.  Here $i$ denotes the tag mode, and $\varepsilon_{\rm ST}$
  and $\varepsilon_{\rm DT}$ are the efficiencies of
  selecting the ST and DT candidates, respectively.

  The same selection criteria as reported in
  Refs.~\cite{epjc76,cpc40,bes3-pimuv,bes3-Dp-K1ev,bes3-etaetapi} are
  used in this analysis.  Charged tracks are required to have polar
  angle ($\theta$) within $|\cos\theta|<0.93$, and except for
  those from $K^0_S$ decays, are required to originate from an
  interaction region defined by $|V_{xy}|<1$~cm and $|V_z|<10$~cm,
  where $|V_{xy}|$ and $|V_z|$ refer to the distances of closest
  approach of the reconstructed track to the interaction point in
  the $xy$ plane and the $z$ direction (along the beam), respectively.

  Particle identification (PID) of  charged kaons and pions is implemented with
  the $dE/dx$  and TOF information. For muon identification,
  the EMC  information is also  included. For each charged  track, the
  combined confidence levels for the electron, muon, pion, and
  kaon  hypotheses~($CL_e$,   $CL_\mu$,  $CL_\pi$,  and   $CL_K$)  are
  calculated.  The charged tracks satisfying $CL_{K(\pi)}>CL_{\pi(K)}$
  are identified  as kaon (pion)  candidates. The muon  candidates are
  required   to  satisfy   $CL_{\mu}>$  0.001,   $CL_{\mu}>CL_e$,  and
  $CL_{\mu}>CL_K$, and  their deposited energy in the  EMC is required
  to   be   within   (0.15,\,0.25)\,GeV   to   suppress   backgrounds
  misidentified from charged hadrons.

  The $K_S^0$ candidates are selected from pairs of opposite charged
  tracks with $|V_{z}|<20$~cm, but without requirements on $|V_{xy}|$.
  The two tracks are designated as pions without PID requirements,
  constrained to a common vertex and required to have an invariant
  mass satisfying $|M_{\pi^+\pi^-} - m_{K_S^0}| < 12~{\rm MeV}/c^2$,
  where $m_{K_S^0}$ is the $K_S^0$ nominal mass~\cite{pdg2018}.  The selected $K_S^0$
  candidate must have a decay length greater than two times the vertex
  resolution.

  Photon candidates are selected using EMC information. It is required
  that the shower time is within 700~ns of the event start time, the
  shower energy must be greater than 25 (50)~MeV in the barrel~(end-cap)
  region~\cite{BESCol}, and the opening angle between the
  candidate shower and any charged tracks must be greater than
  $10^{\circ}$.

  The $\pi^0$ candidates are selected from photon pairs with invariant
  mass within $(0.115,\,0.150)$\,GeV$/c^{2}$.  To improve the momentum
  resolution, a one constraint~(1-C) kinematic fit is performed
  constraining the pair's $\gamma\gamma$ invariant mass to the
  $\pi^{0}$ nominal mass~\cite{pdg2018}, and the $\chi^2_{\rm 1-C}$ of
  the 1-C (mass-constraint) kinematic fit is required to be less than
  200.

  The energy difference ($\Delta E$) and beam-constrained mass
  ($M_{\rm BC}$) are used to select ST $D^-$ candidates, where
\begin{equation}
\Delta E\equiv E_{D^-}-E_{\mathrm{beam}}
\end{equation}
and
\begin{equation}
M_{\rm BC}\equiv\sqrt{E_{\mathrm{beam}}^{2}-|\vec{p}_{D^-}|^{2}}.
\end{equation}
\noindent $E_{\rm beam}$ is the beam energy, and $\vec{\mkern1mu
  p}_{D^-}$ and $E_{D^-}$ are the total momentum and energy of the ST
candidate calculated in the $e^+e^-$ rest frame, respectively. The $D^-$ candidates
are expected to concentrate around zero in the $\Delta E$ distribution
and around the nominal $D^-$ mass in the $M_{\rm BC}$
distribution. For each tag mode, the one with minimum $|\Delta E|$ is
retained. Combinatorial backgrounds in the $M_{\rm BC}$ distributions
are suppressed with a requirement of $\Delta E\in (-0.055,0.045)$~GeV
for tags containing $\pi^0$ and $\Delta E\in (-0.025,0.025)$~GeV for
other tags.

\begin{figure}[htbp]\centering
\includegraphics[width=1.0\linewidth]{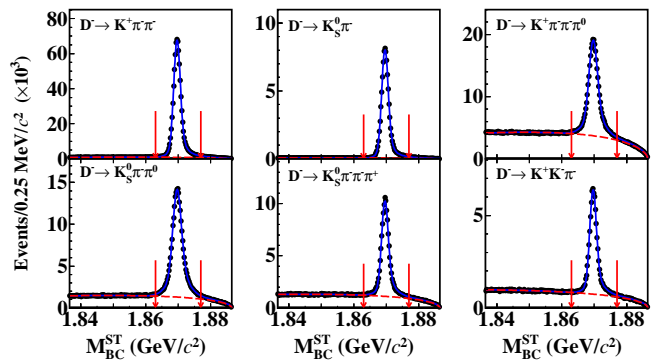}
\caption{
Fits to the $M_{\rm BC}$ distributions of the ST candidate events.
The dots with error bars are data,
the blue solid curves are the fit results,
the red dashed curves are the fitted backgrounds, and
the pair of red arrows in each sub-figure denote the ST $D^-$ signal region.
}\label{fig:datafit_Massbc}
\end{figure}

For each tag mode, the ST yield is determined by fitting the $M_{\rm
  BC}$ distribution of the candidates surviving all above
requirements. In the fit, the $D^-$ signal is modeled with a shape
obtained from an MC simulation convolved with a double Gaussian describing the
difference between data and MC simulations, and the
combinatorial background is described by an ARGUS
function~\cite{argus}. The resulting fits to the $M_{\rm BC}$
distributions for each mode are shown in
Fig.~\ref{fig:datafit_Massbc}. Candidates in the $M_{\rm BC}$
signal region, $(1.863,1.877)$ GeV$/c^2$, are kept for further
analysis. The ST yields in data and the ST efficiencies for individual
tags are shown in Table~\ref{tab:styields}.  Summing over the ST
yields for all tags gives a total yield of $N^{\rm tot}_{\rm
  ST}=1522474\pm2215$, where the uncertainty is statistical.

\begin{table}[htbp]\centering
  \caption{
    Summary of ST yields ~($N^{i}_{\rm ST}$), ST efficiencies ~($\varepsilon^{i}_{\rm ST}$) and DT efficiencies ~($\varepsilon^{i}_{\rm DT}$) for different tag modes. Uncertainties are statistical only. Efficiencies do not include the BFs of $K^0_S\to\pi^+\pi^-$, $\pi^0\to\gamma\gamma$, and $\omega \to\pi^+\pi^-\pi^0$.}
\label{tab:styields}
\small
\begin{tabular}{lcccc}
\hline
  Tag mode & $N^i_{\rm ST}$ & $\epsilon^i_{\rm ST}$ (\%) &
  $\epsilon^{i}_{\rm DT}$ (\%) \\
  \hline
  $D^-\to K^{+}\pi^{-}\pi^{-}$            & $782669\pm\hspace{0.15cm}990$  & $50.61\pm0.06$          &$4.28\pm0.05$ \\
  $D^-\to K^0_{S}\pi^{-}$               & $\hspace{0.15cm}91345\pm\hspace{0.15cm}320$&$50.41\pm0.17$ &$4.57\pm0.06$ \\
  $D^-\to K^{+}\pi^{-}\pi^{-}\pi^{0}$     & $251008\pm1135$ & $26.74\pm0.09$                  & $1.89\pm0.04$  \\
  $D^-\to K^0_{S}\pi^{-}\pi^{0}$        & $215364\pm1238$ & $27.29\pm0.07$                    & $2.26\pm0.06$  \\
  $D^-\to K^0_{S}\pi^{+}\pi^{-}\pi^{-}$ & $113054\pm\hspace{0.15cm}889$ &$28.29\pm0.12$       & $2.16\pm0.09$ \\
  $D^-\to K^{+}K^{-}\pi^{-}$     & $\hspace{0.15cm}69034\pm\hspace{0.15cm}460$&$40.87\pm0.24$ & $3.05\pm0.05$ \\
\hline
\end{tabular}
\end{table}

The $D^+\to \omega\mu^+\nu_\mu$ candidates are selected from the
remaining charged tracks and photons that have not been used for the
ST reconstruction. Each candidate must have three good charged tracks
and one $\pi^0$ candidate. If there are multiple neutral pions, the
one with the minimum $\chi^2_{\rm 1-C}$ is chosen. One of the three
charged tracks must be identified as a muon, and the other two as
$\pi^+\pi^-$. The total charge of the DT event is required to be
zero. The $\omega$ candidates are selected from $\pi^+\pi^-\pi^0$
combinations, and we require $|M_{\pi^+\pi^-\pi^0}-m_\omega|<0.025$
GeV/$c^2$, where $m_\omega$ is the $\omega$ nominal
mass~\cite{pdg2018} and $M_{\pi^+\pi^-\pi^0}$ is the invariant mass of
the $\pi^+\pi^-\pi^0$ combination. If two $\pi^+\pi^-\pi^0$
combinations can be formed due to mis-identification between $\pi^+$
and $\mu^+$, the one with $M_{\pi^+\pi^-\pi^0}$  closer to $m_\omega$
is kept as the $\omega$ candidate.  To suppress backgrounds from the
SL decays $D^+\to \bar K^*(892)^0\mu^+\nu_\mu$ with $\bar
K^*(892)^0\to K^0_S (\pi^+\pi^-)\pi^0$, we require
$|M_{\pi^+\pi^-}-m_{K^0_S}|>0.015$~GeV/$c^2$ and
$|M_{\pi^+_{\mu\to\pi}\pi^-}-m_{K^0_S}|>0.015$~GeV/$c^2$, where
$M_{\pi^+\pi^-}$ and $M_{\pi^+_{\mu\to\pi}\pi^-}$ are the invariant masses of the $\pi^+\pi^-$ and $\mu^+\pi^-$ combinations, respectively;
and $\pi^+_{\mu\to\pi}$
denotes that the mass of the muon candidate has been replaced by the $\pi^+$
mass.
These requirements correspond to approximately four times the fitted mass resolution of $K^0_S$ around its nominal mass. To suppress backgrounds from the hadronic
decays $D^+\to K^0_S(\pi^0\pi^0)\pi^+\pi^+\pi^-$,
the invariant mass of the system recoiling against the $D^-\pi^+_{\mu\to\pi}\pi^+\pi^-$ combination ($M^{\rm
  recoil}_{D^-\pi^+_{\mu\to\pi}\pi^+\pi^-}$) is required to be outside
the range of (0.45, 0.55) GeV/$c^2$. The peaking backgrounds
from the hadronic decays $D^+\to\omega\pi^+$ and
$D^+\to\omega\pi^+\pi^0$ are suppressed by requiring
$M_{\omega\mu^+}<1.5$ GeV/$c^2$ and $E_{\rm extra~\gamma}^{\rm
  max}<0.15$ GeV. Here, $M_{\omega\mu^+}$ is the invariant mass of the
$\omega\mu^+$ combination and $E_{\rm extra~\gamma}^{\rm max}$ is the
maximum energy of any photon that is not used in the DT selection.

The neutrino of the SL $D$ decay is undetectable by the BESIII
detector. The information of the $D^+\to\omega\mu^+\nu_\mu$ decay is
inferred by the difference between the missing energy
($E_{\mathrm{miss}}$) and the missing momentum
($|\vec{p}_{\mathrm{miss}}|$) of the observed particles of the DT
event calculated in the $e^+e^-$ center-of-mass frame,
$U_{\mathrm{miss}}\equiv
E_{\mathrm{miss}}-|\vec{p}_{\mathrm{miss}}|$. Here,
$E_{\mathrm{miss}}\equiv E_{\mathrm{beam}}-E_{\omega}-E_{\mu^{+}}$ and
$\vec{p}_{\mathrm{miss}}\equiv\vec{p}_{D^+}-\vec{p}_{\omega}-\vec{p}_{\mu^{+}}$,
where $E_{\omega\,(\mu^+)}$ and $\vec{p}_{\omega\,(\mu^+)}$ are the
energy and momentum of the $\omega$\,($\mu^+$) candidates, respectively. The
$U_{\mathrm{miss}}$ resolution is improved by constraining the $D^+$
energy and momentum with the beam energy and $\vec{p}_{D^+}$ =
$-\hat{p}_{D^-}\sqrt{E_{\mathrm{beam}}^{2}-m_{D^-}^{2}}$, where
$\hat{p}_{D^-}$ is the unit vector in the momentum direction of the
tagged $D^-$ and $m_{D^-}$ is the $D^-$ nominal mass~\cite{pdg2018}.

\begin{figure}[htbp] \centering
  \vspace*{-0.2cm}
  \includegraphics[width=0.95\linewidth]{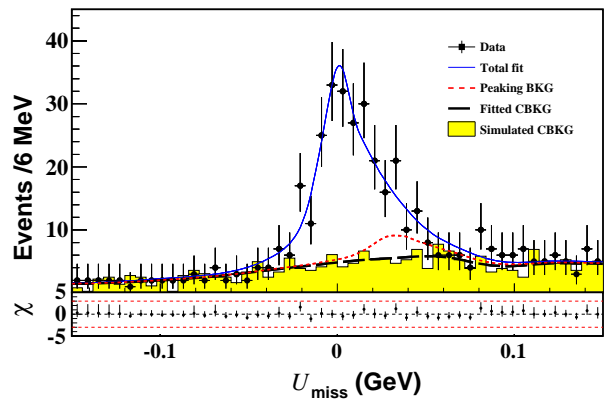}
  \vspace*{-0.5cm}
\caption{The results of a fit to the $U_{\rm miss}$ distribution of the
$D^{+}\rightarrow\omega\mu^{+}\nu_{\mu}$ candidate events.
The dots with error bars are data and
the blue solid curve is the fit result.
The yellow hatched histogram is the MC-simulated combinatorial background (Simulated CBKG),
the black dashed curve is the result of a fit of the combinatorial background (Fitted CBKG),
and the difference between the red dotted and black dashed curves is the peaking background
of $D^+\to \omega\pi^+\pi^0$ (Peaking BKG). The bottom plot shows the $\chi$ distribution
obtained from the fit.}
\label{fig:fit_Umistry1}
\end{figure}

The $U_{\mathrm{miss}}$ distribution of the accepted DT events of data
is shown in Fig.~\ref{fig:fit_Umistry1}. An unbinned {\rm maximum likelihood} fit to this
distribution is used to determine the SL decay yield. The
shapes of all the components in the fit are obtained from MC
simulations, including the SL signal, the peaking background from the
hadronic decays $D^+\to \omega\pi^{+}\pi^{0}$, and other backgrounds,
while their yields are left free.  The number of
$D^+\to\omega\mu^+\nu_\mu$ decays obtained is $N_{\rm
  DT}=194\,\pm\,20$, where the uncertainty is statistical.

The fourth column of Table~\ref{tab:styields} lists the DT
efficiencies for individual tag modes. The signal efficiency weighted by
the ST yields in data is $\varepsilon_{\rm SL}=(8.15\,\pm\,0.07)\%$.
Detailed studies show that the momentum and $\cos\theta$ distributions
of $\omega$ and $\mu^+$ of data are modeled well by MC simulations.
The BF of the $D^+\to\omega \mu^+\nu_\mu$ decay is obtained by Eq.~(\ref{eq:bf}) to be
\begin{equation}
{\mathcal B}_{D^+\to\omega \mu^+\nu_\mu}=(17.7\pm1.8\pm1.1)\times 10^{-4}, \nonumber
\end{equation}
where the first uncertainty is statistical and the second systematic.

With the DT method, most systematic uncertainties arising from the ST
side cancel. In the BF measurement, the systematic uncertainties arise
from the following sources. The uncertainty in the total ST yield,
which is mainly from the uncertainty due to the $M_{\rm BC}$ fit of
the ST candidates, has been studied in
Refs.~\cite{epjc76,cpc40,bes3-pimuv} and is assigned as 0.5\%. The
tracking and PID  efficiencies of the pion and muon are studied by
analyzing the DT hadronic $D\bar D$ events and
$e^+e^-\to\gamma\mu^+\mu^-$ events, respectively. The systematic
uncertainties associated with the pion tracking (PID), muon tracking
(PID) are assigned to be 0.2\% (0.3\%) and 0.3\% (0.3\%),
respectively.  The $\pi^0$ efficiency, including effects of photon
selection, the 1-C kinematic fit, and the mass window, is studied with
DT hadronic $D\bar D$ decays~\cite{epjc76,cpc40}, and a systematic
uncertainty of 0.7\% is assigned to each $\pi^0$. The uncertainty of
the $E_{\rm extra~\gamma}^{\rm max}$ requirement is estimated to be
4.4\% by analyzing the DT $D\bar D$ events of $D^+\to
\omega(\pi^{+}\pi^{-}\pi^{0}) e^+ \nu_{e}$, $D^+\to
K^0_S(\pi^+\pi^-)\pi^{0} e^+ \nu_{e}$, $D^+\to K^0_S(\pi^+\pi^-) e^+
\nu_{e}$, and $D^+\to K^0_S(\pi^+\pi^-) \pi^{+} \pi^{0}$.  The
uncertainties due to the $M_{\omega\mu^+}$ requirement and the $K^0_S$
rejection ($M_{\pi^+\pi^-}$, $M_{\pi^+_{\mu\to\pi}\pi^-}$, and $M^{\rm
  recoil}_{D^-\pi^+_{\mu\to\pi}\pi^+\pi^-}$) are evaluated by
repeating measurements varying the nominal requirements by $\pm
\,0.05$ GeV/$c^2$ and $\pm \,0.005$~GeV$/c^2$, respectively, and they
are found to be negligible. The uncertainty originating from the
$U_{\rm miss}$ fit is assigned to be 3.4\%, which is estimated with
alternative fit ranges and signal and background shapes. The uncertainty
due to the limited MC size is 0.5\%. The uncertainty in the MC model
is assigned to be 2.3\%, by comparing our nominal DT efficiency with
one obtained using an ISGW model~\cite{isgw2}. All these systematic
uncertainties are assumed to be independent, and their quadratic sum
gives a total systematic uncertainty of 6.3\%.

To summarize, by analyzing the data sample with an integrated
luminosity of 2.93 fb$^{-1}$ collected at $\sqrt{s}=3.773$ GeV with
the BESIII detector, we report the first observation and BF measurement of the SL decay $D^+\to
\omega\mu^+\nu_\mu$. Table~\ref{table:comparison}
shows the comparison of our BF to various theoretical calculations of
$D^+\to \omega \mu^+ \nu_{\mu}$ decay. Our BF is consistent with
the predicted values based on the LFQM, CCQM, and LCSR
methods~\cite{cheng,soni1,soni2,lcsr2}, but differs from those based on the
$\chi$UA~\cite{Sekihara} and RQM~\cite{rqm}  methods by $2.5\sigma$ and $1.5\sigma$, respectively.
Combining the ${\mathcal B}_{D^+\to\omega\mu^+\nu_\mu}$
measured in this work with the world average ${\mathcal B}^{\rm
  PDG}_{D^+\to\omega e^+\nu_e}=(16.9\pm1.1)\times
10^{-4}$~\cite{CLEO-omegaev,BESIII-omegaev,pdg2018}, we obtain the BF ratio to be ${\mathcal
  B}_{D^+\to \omega \mu^+\nu_\mu}/{\mathcal B}^{\rm PDG}_{D^+\to
  \omega e^+\nu_e}=1.05\pm0.14$. It agrees with the standard model
predictions~(0.93-0.99) ~\cite{cheng,Sekihara,soni1,soni2,rqm} within
uncertainties,
and implies no violation of $e$-$\mu$ LFU found with current statistics.

\begin{table*}[htpb]
\centering
\caption{Comparison of the BFs between $D^+\to\omega e^+\nu_e$ and $D^+\to \omega \mu^+ \nu_{\mu}$.}
\label{table:comparison}
\small
\begin{tabular}{lccccccc}
  \hline
     & CCQM \cite{cheng} & $\chi$UA~\cite{Sekihara} & LFQM \cite{soni1} &LCSR \cite{soni2} & LCSR \cite{lcsr2}  &RQM \cite{rqm}& Measurement      \\
\hline
${\mathcal B}_{D^+\to \omega \mu^+ \nu_{\mu}}$ ($\times 10^{-4}$)& 17.8& 22.9 & $20\pm2$ &$18.5^{+1.9}_{-1.3}$ &$17.3^{+4.8}_{-4.0}$& 20.8  &$17.7\pm1.8\pm1.1$ \\ \hline
${\mathcal B}_{D^+\to \omega e^+ \nu_{e}}$    ($\times 10^{-4}$) & 18.5& 24.6 & $21\pm2$ &$19.3^{+2.0}_{-1.4}$  &$17.4^{+4.8}_{-4.0}$& 21.7 &$16.9\pm1.1$~\cite{pdg2018} \\  \hline
${\mathcal B}_{D^+\to \omega \mu^+\nu_\mu}/{\mathcal B}^{\rm PDG}_{D^+\to \omega e^+\nu_e}$ & 0.96& 0.93 & $0.95$ & 0.96&0.99& $0.96$ &$1.05\pm0.14$ \\   \hline
\end{tabular}
\end{table*}

The BESIII collaboration thanks the staff of BEPCII and the IHEP computing center for their strong support. This work is supported in part by National Key Basic Research Program of China under Contract No. 2015CB856700; National Natural Science Foundation of China (NSFC) under Contracts Nos. 11675200, 11625523, 11635010, 11735014, 11822506, 11835012, 11961141012; the Chinese Academy of Sciences (CAS) Large-Scale Scientific Facility Program; Joint Large-Scale Scientific Facility Funds of the NSFC and CAS under Contracts Nos. U1632109, U1532257, U1532258, U1732263, U1832207; CAS Key Research Program of Frontier Sciences under Contracts Nos. QYZDJ-SSW-SLH003, QYZDJ-SSW-SLH040; 100 Talents Program of CAS; INPAC and Shanghai Key Laboratory for Particle Physics and Cosmology; ERC under Contract No. 758462; German Research Foundation DFG under Contracts Nos. Collaborative Research Center CRC 1044, FOR 2359; Istituto Nazionale di Fisica Nucleare, Italy; Ministry of Development of Turkey under Contract No. DPT2006K-120470; National Science and Technology fund; STFC (United Kingdom); The Knut and Alice Wallenberg Foundation (Sweden) under Contract No. 2016.0157; The Royal Society, UK under Contracts Nos. DH140054, DH160214; The Swedish Research Council; U. S. Department of Energy under Contracts Nos. DE-FG02-05ER41374, DE-SC-0010118, DE-SC-0012069.

\end{document}

%% file: authorlist_CWR_27Nov2019.tex
\author{
M.~Ablikim$^{1}$, M.~N.~Achasov$^{10,e}$, P.~Adlarson$^{64}$, S. ~Ahmed$^{15}$, M.~Albrecht$^{4}$, A.~Amoroso$^{63A,63C}$, Q.~An$^{60,48}$, ~Anita$^{21}$, Y.~Bai$^{47}$, O.~Bakina$^{29}$
% ...
\\
\vspace{0.2cm}
(BESIII Collaboration)\\
\vspace{0.2cm} {\it
$^{1}$ Institute of High Energy Physics, Beijing 100049, People's Republic of China\\
% ...
$^{4}$ Bochum Ruhr-University, D-44780 Bochum, Germany\\
% ...
$^{10}$ G.I. Budker Institute of Nuclear Physics SB RAS (BINP), Novosibirsk 630090, Russia\\
% ...
$^{15}$ Helmholtz Institute Mainz, Johann-Joachim-Becher-Weg 45, D-55099 Mainz, Germany\\
% ...
$^{21}$ Indian Institute of Technology Madras, Chennai 600036, India\\
% ...
$^{29}$ Joint Institute for Nuclear Research, 141980 Dubna, Moscow region, Russia\\
% ...
$^{47}$ Southeast University, Nanjing 211100, People's Republic of China\\
$^{48}$ State Key Laboratory of Particle Detection and Electronics, Beijing 100049, Hefei 230026, People's Republic of China\\
% ...
$^{60}$ University of Science and Technology of China, Hefei 230026, People's Republic of China\\
% ...
$^{63}$ (A)University of Turin, I-10125, Turin, Italy; (B)University of Eastern Piedmont, I-15121, Alessandria, Italy; (C)INFN, I-10125, Turin, Italy\\
$^{64}$ Uppsala University, Box 516, SE-75120 Uppsala, Sweden\\
% ...
$^{e}$ Also at the Novosibirsk State University, Novosibirsk, 630090, Russia\\
% ...
}
}